\documentstyle[preprint,aps]{revtex}

\begin{document}
\draft
\title{Edge and bulk effects in the Terahertz-photoconductivity of an antidot 
superlattice}
\author{B. G. L. Jager, S. Wimmer, A. Lorke, and J. P. Kotthaus}
\address{Sektion Physik and Center for NanoScience, 
Ludwig-Maximilians-Universit\"at M\"unchen, Geschwister-Scholl-Platz 1, 
D-80539 M\"unchen, Germany} 
\author{W. Wegscheider\cite{AddWW} and M. Bichler}
\address{Walter Schottky Institut, TU M\"unchen, Am Coulombwall 
D-85748 Garching, Germany}

\date{\today}
\maketitle
%---------------------------------------------------------------------

\begin{abstract}

We investigate the Terahertz(THz)-response of a square  antidot superlattice 
by means of photoconductivity measurements  using a 
Fourier-transform-spectrometer. We detect, spectrally  resolved, the cyclotron 
resonance and the fundamental  magnetoplasmon mode of the periodic 
superlattice. In the  dissipative transport regime both resonances are 
observed in the  photoresponse. In the adiabatic transport regime, at integer 
filling  factor $\nu =2$, only the cyclotron resonance is observed. From  this 
we infer that different mechanisms contribute to converting the  absorption of 
THz-radiation into photoconductivity in the cyclotron and in the 
magnetoplasmon resonances, respectively. In the dissipative transport regime,  
heating of the electrons via resonant absorption of the THz-radiation  in the 
two-dimensional (2D) bulk is the main mechanism of photoconductivity in both  
resonances. In the case of the cyclotron resonance, and especially in the  
adiabatic transport regime, we find an additional contribution to  
photoconductivity which we interprete as being caused by  
THz-absorption-induced backscattering of edge states. Since the  
characteristic length over which the magnetoplasmon oscillator strength 
decreases from the 2D bulk of the superlattice towards its  edges is by about 
an order of magnitude larger than the  extent of the edge-states into the 2D 
bulk, the  magnetoplasmon is not  able to induce such backscattering of 
edge-states.  Thus in the adiabatic transport regime, when only the 
edge-states  contribute to electric conduction, THz-absorption in the 
magnetoplasmon cannot be converted into photoconductivity.
%---------------------------------------------------------------------
\end{abstract}

%---------------------------------------------------------------------

\section{Introduction}

The magnetotransport properties of a two-dimensional electron gas (2DEG) in 
high magnetic fields are successfully described in the edge channel or 
Landauer-B\"uttiker picture 
\cite{Buttiker}\cite{Landauer}\cite{Streda}\cite{Jain}. According to this 
model the spatial current distribution in a 2DEG in high magnetic fields is 
strongly influenced by the filling factor of the 2DEG. At the edges of a Hall 
bar the Landau levels (LLs) of the unbounded 2DEG are bent upwards in energy 
due to the confining potential. The intersections of these upwards bent Landau 
levels with the Fermi energy form the so-called edge states \cite{Laughlin 
81}\cite{Halperin 82}\cite{CSG 92}. Electrons in edge states at opposite edges 
of the Hall bar flow into opposite directions. At integer filling factors the 
Fermi energy lies in the regime of localized states between two Landau levels 
of the 2D-bulk, and a sufficiently small electric current is carried by the 
edge states only. In thermal equilibrium all edge states at the same sample 
edge have the same chemical potential. Scattering between edge states of the 
same sample edge (inter-LL scattering) is likely to occur after a distance 
which is called the equilibration length $l_{eq}$. However, this kind of 
scattering event does not lead to a nonzero magnetoresistance $\rho_{xx}$ 
along the current direction as it cannot reverse an electron's direction of 
motion. Over distances up to the equilibration length a nonequilibrium 
population between edge states (of the same edge) can be maintained. The 
transport regime in which no equilibration between edge states (of the same 
edge) occurs is called the adiabatic regime. At noninteger filling factors the 
Fermi energy lies within the topmost partially filled Landau level. This 
topmost (further also referred to as Nth) Landau level can form a bulk 
conductance channel that allows backscattering of electrons, which leads to a 
nonzero longitudinal magnetoresistance $\rho_{xx}$ \cite{McEuen} 
\cite{Komiyama}. The Nth channel is almost perfectly decoupled from the lower 
N$-1$ edge states \cite{McEuen}. This means that the edge states propagate 
unaffected by the bulk channel just as they would do for integer filling 
factors. 

Here we study the photoconductivity of nanostructured 2DEGs caused by 
intraband absorption in the quantum Hall regime. Intraband photoconductivity 
of 2DEGs \cite{Shiraki} has successfully been described by Neppl {\em et al.} 
\cite{Neppl} using a bolometric model, which was originally employed to 
explain the mechanism of photoconductivity caused by intersubband resonance in 
2DEGs. Later the same model was also applied to cyclotron-resonance-induced 
photoconductivity in the quantum Hall regime \cite{CRPhoto} and to 
photoconductivity of systems with reduced dimensionality \cite{NanoPhoto}. In 
cyclotron resonance, the absorption of radiation excites single electrons from 
the N$-1$st (bulk) LL below the Fermi energy to the Nth LL above the Fermi 
energy. Within a bolometric model the excited electrons thermalize via 
electron-electron interactions and a new quasi-equilibrium corresponding to a 
higher electron temperature is established. As the longitudinal resistance 
$\rho_{xx}$ depends on temperature, the increased electron temperature results 
in a change of the longitudinal resistance, $\Delta \rho_{xx}$, which is 
measured in a photoconductivity experiment. 

In more recent photoconductivity experiments, an enhanced cyclotron resonance 
amplitude in the adiabatic transport regime close to a current injecting 
contact \cite{Merz} was observed. The longitudinal resistance in the adiabatic 
regime at low temperatures and bias currents under illumination by a 
far-infrared laser was observed to show a sharp maximum \cite{Diessel} at the 
cyclotron resonance. The authors concluded that the cyclotron resonance 
absorption process induces a nonequilibrium population of edge states. They 
infer that this also leads to inter-edge scattering (i.e. to scattering events 
that are able to reverse the electrons' direction of motion) and thus to an 
increase in longitudinal resistance. This increase in resistance can be 
observed directly in photoconductivity \cite{Merz} and in magnetotransport 
\cite{Diessel}.

We investigate the photoconductivity properties of an antidot superlattice 
which allows us to excite the cyclotron resonance and the fundamental 
magnetoplasmon mode by illumination with THz-radiation. We find that at 
noninteger filling factors mainly resonant electron heating contributes to 
photoconductivity in both absorption processes. Under adiabatic transport 
conditions at a filling factor of $\nu =2$, on the contrary, only the 
cyclotron resonance (CR) is observed. Here, in addition, the CR amplitude has 
a different dependence on the applied bias current than the magnetoplasmon and 
the cyclotron resonance amplitudes in the dissipative transport regime. This 
indicates that the photoresponse in the adiabatic transport regime is 
generated by a process different from electron heating which is mainly 
responsible for the photosignal in the dissipative regime. 

\section{Sample details and experimental technique}

The sample investigated here is a square antidot superlattice with a period of 
500~nm, fully covering the distance of $39~\mu m$ between adjacent voltage 
probes on a $39~\mu m$ wide Hall bar (Fig.~\ref{ADOTS}).

The antidot-superlattice is prepared on a GaAs/AlGaAs heterostructure with a 
mobility (at $T=4.2$ K) of $\mu = 1.53\cdot 10^{5}~cm^2/Vs$ and an electron 
density $n_{s}= 3.45\cdot 10^{11}cm^{-2}$ in the unpatterned 2DEG. The 2DEG 
has a distance of 37~nm from the sample surface. The antidots are written by 
e-beam lithography and transferred into the 2DEG by shallow wet etching. In 
the patterned region the density is smaller by about 7$\% $ compared to the 
unpatterned 2DEG. 

The antidots have a triangular shape (width~=~300~nm, height~=~200~nm) 
\cite{Jager}. However, this is not essential for the results presented here. 

We study the photoresponse, in particular the change in resistance $\Delta 
R_{xx} =\Delta U_{xx} /I$ in response to the THz-illumination, by measuring 
the photo-induced voltage $\Delta U_{xx} $ between two voltage probes in 
four-point geometry as shown in Fig.~\ref{ADOTS}. We apply dc bias currents 
$I$ of up to 10~$\mu A$ between source and drain contacts. The bath 
temperature during the measurements is 4.2~K, and magnetic fields of up to 
12~T are applied perpendicular to the plane of the 2DEG. 

We use a broad-band mercury lamp to irradiate our samples. The light source is 
modulated by a Fourier-transform-spectrometer to spectrally analyze the 
measured photoresponse $\Delta U_{xx} $. In the spectral range of interest to 
us, between 0 and $200~cm^{-1}$ (0 to 6 THz), the Hg lamp has an integrated 
intensity of a few micro-Watts. 
%---------------------------------------------------------------------

\section{Experimental results}

In Fig.~\ref{02131484T}(a) typical data in the dissipative regime at a 
magnetic field of $8.4~T$ ($\nu \approx 1.6$) and a bias current of $1~\mu A$ 
applied from the source to the drain contact are shown. As can be seen there, 
we detect the cyclotron resonance (CR) and a magnetoplasmon resonance (MP). 
The photoresponse $\Delta U_{xx}$ was recorded between contacts 3 and 4. 
Fig.~\ref{02131484T}(b) shows a spectrum taken at the reference section on the 
same Hall bar (contacts 1 and 2). There only the cyclotron resonance is 
resolved. 

We record the photoresponse spectra for a series of magnetic fields between 0 
and 12~T and fit the spectral resonance position of the magnetoplasmon as a 
function of the applied magnetic field by the formula 
$\omega^2=\omega_{0}^2+\omega_{c}^2$ \cite{Chaplik}\cite{Theis} to extrapolate 
the zero-magnetic-field plasmon frequency $\omega_{0}$. This $\omega_{0}$ is 
compared to the calculated frequency $\omega_{P}$ of a plasmon in a 
two-dimensional electron gas with a two-dimensional modulation of the charge 
density with periods $a_{x}$ and $a_{y}$ along the $x-$ and $y-$ directions, 
respectively. We determine $\omega_{P}$ from sample-specific parameters via 
the formula \cite{Stern 67}\cite{Eliasson 86}

\begin{equation}
\label{Plasmon}
\omega_P^2 = \frac{N_se^2}{2m^*\epsilon_{eff}(k)\epsilon_0}\cdot \sqrt{(n_{x} 
\cdot \frac{2\pi}{a_{x}})^2+(n_{y} \cdot \frac{2\pi}{a_{y}})^2} =: 
\frac{N_se^2}{2m^*\epsilon_{eff}(k)\epsilon_0}\cdot k, 
\end{equation}

where in our case $a_{x}=a_{y}=a=500nm$. $N_{s}$ is the 2D electron density, 
$m^{*}$ the effective mass, $k$ the wave-vector as defined by the above 
formula and $\epsilon_{eff}$ the effective dielectric constant. For 
$\epsilon_{eff}$ we use\cite{Chaplik} $\epsilon_{eff}(k) = \epsilon_{GaAs} / 
(1+\frac{\epsilon -1}{\epsilon +1} e^{-2kd})$,where the distance $d$ of the 
2DEG from the sample surface equals $d=37nm$ in our samples. The effective 
mass $m^{*}=0.07$ is deduced from the frequency of the cyclotron resonance, 
measured in the reference section. We fit our data by the 
$(n_{x},n_{y})=(1,0)$ mode and obtain very good agreement between theory and 
experiment. Therefore, the resonance beside the cyclotron resonance is 
identified as the fundamental (1,0) or the degenerate (0,1) magnetoplasmon 
mode in the superlattice.

Figure \ref{031780T}(a) shows photoconductivity spectra taken at a magnetic 
field of B~=~8.0~T (corresponding to a filling factor of $\nu =1.7$ of the 
2DEG) for bias currents applied between source and drain contacts ranging from 
$+10~\mu A$ to $-10~\mu A$ in steps of $2~\mu A$. As expected, with no bias 
current applied no photoconductivity is observed (curve $I=0$). For bias 
currents of $I=\pm 2~\mu A$ both the cyclotron resonance (CR) and the 
magnetoplasmon (MP) are observed in the photoresponse $\Delta U_{xx}$ and have 
a large amplitude already. For higher bias currents up to $I=\pm 10~\mu A$ the 
photoresponse amplitude in the cyclotron resonance slightly increases and 
finally decreases again. The photoresponse amplitude in the magnetoplasmon 
absorption, on the other hand, decreases monotonically.

Fig.~\ref{031780T}(b) shows data analogous to those in Fig.~\ref{031780T}(a), 
but for a magnetic field of $B=6.8~T$, corresponding to a filling factor of 
$\nu = 2$. Under these conditions, up to high bias-currents of $I=\pm 8~\mu A$ 
only the cyclotron resonance is observed, but not the magnetoplasmon. Only for 
very high bias currents of $I=\pm 10~\mu A$ a magnetoplasmon-induced 
photoresponse is beginning to develop. The amplitude of the cyclotron 
resonance increases linearly with increasing bias current. 

For a filling factor of $\nu =1.9$ (B~=~7.2~T) (not shown) both the cyclotron 
resonance and the magnetoplasmon are observed in the photoresponse, but the 
magnetoplasmon only for bias currents higher than about $I=\pm 3~\mu A$. 

Summarizing, at integer filling factor $\nu =2$ of the 2DEG only the cyclotron 
resonance is observed in photoresponse. At $\nu <2$ both the magnetoplasmon 
and the cyclotron resonance are observed in photoresponse. 

Figure \ref{0317UI} shows the amplitudes of the spectra from 
Figs.~\ref{031780T}(a) and \ref{031780T}(b) as a function of the bias current 
between source and drain. Open circles correspond to the cyclotron resonance 
maximum, black dots to the magnetoplasmon signal. The dashed interpolating 
lines are guides to the eye only. For comparison, the differential 
longitudinal resistance $dU_{xx}/dI_{bias}$ as a function of the bias current, 
recorded in the same geometry as the photoresponse, is plotted (solid line). 
%---------------------------------------------------------------------

\section{Discussion}

Figure \ref{0317UI} demonstrates that the amplitude of the magnetoplasmon 
signal qualitatively follows the differential longitudinal resistance 
$dU_{xx}/dI_{bias}$ for different magnetic fields (filling factors). It will 
be shown in the following that therefore the magnetoplasmon signal can be 
explained by heating of the electron gas corresponding to the bolometric model 
\cite{Neppl}. 

In the bolometric model the photoresponse of a sample is regarded as a change 
in the voltage (or resistance or conductivity, depending on whether a voltage, 
resistance or conductivity is measured in the specific sample geometry) caused 
by an increased electron temperature. In our case, the longitudinal voltage is 
measured in a four point geometry (Fig.~\ref{ADOTS}). Thus the photoresponse 
corresponds, in terms of the bolometric model, to a temperature-induced change 
in the longitudinal voltage $\Delta U_{xx} = dU_{xx}/dT \cdot \Delta T$. The 
increase in electron temperature, $\Delta T$, is provided by the resonant 
absorption of the THz radiation. Since the absorption coefficient does not 
strongly depend on magnetic field or filling factor and since filling 
factor-induced oscillations of the electronic specific heat are rather weak, 
we can for simplicity assume that $\Delta T$ does not depend on magnetic field 
or filling factor in the present, small signal limit. Thus, the photoresponse 
is determined by the dependence $dU_{xx}/dT$ of the voltage U on temperature.

On the other hand, a sufficiently high bias current $I$ also produces an 
increase in electron temperature. Thus the differential $dU_{xx}/dI$, is 
directly proportional to $dU_{xx}/dT$. 

The amplitude of the magnetoplasmon-induced photoresponse as a function of 
bias-current $I$ resembles very much the differential longitudinal resistance 
$dU_{xx}/dI$ and thus $dU_{xx}/dT$. Thus the bolometric model can describe the 
photoresponse mechanism at the magnetoplasmon resonance as heating of the 
electron gas. 

A nonzero longitudinal voltage $U_{xx}(T)$ is, on the other hand, a rough 
measure for the amount of conduction through bulk, dissipative electronic 
states. Thus the appearence of the magnetoplasmon in photoconductivity can be 
taken for an indicator of beginning bulk conductivity. 

For $\nu =1.7$ (B~=~8~T) the amplitude of the cyclotron resonance has a 
behaviour similiar to that of the magnetoplasmon (Fig.~\ref{0317UI}(a)), even 
though the features occur at higher bias currents. For $\nu=2$ or B~=~6.8~T 
(Fig.~\ref{0317UI}(c)), in contrast, the amplitude of the cyclotron resonance 
signal increases linearly with increasing bias current while the amplitude of 
the magnetoplasmon response and the longitudinal resistance (solid line) stay 
almost zero up to bias currents of $I=\pm 8~\mu A$. The suppression of the 
magnetoplasmon and the vanishing longitudinal resistance indicate that the 
2D-bulk is still insulating. Thus only the edge-states contribute to the 
conductance at an external magnetic field of B~=~6.8~T ($\nu =2$). Therefore, 
we attribute the cyclotron resonance part of the photoresponse at $\nu =2$ to 
backscattering of the topmost edge states. At B~=~8~T probably both mechanisms 
are important, as will be discussed below.

Summarizing the discussion, the magnetoplasmon absorption leads to a 
photoresponse only in the dissipative transport regime, when the 2D-bulk is 
conducting, which also manifests itself in a nonzero (differential) 
longitudinal magnetoresistance $\rho_{xx} =dU_{xx}/dI$. The cyclotron 
resonance absorption leads to a photoresponse also in the adiabatic transport 
regime, when only the edge-states contribute to electric conductance. 

As for the mechanism of backscattering we follow the argumentation of Diessel 
{\em et al.} and propose that the absorption of THz radiation at the cyclotron 
resonance in the regime of the QHE at $\nu \approx 2$ leads to a 
nonequilibrium population of edge states and thus to an enhanced inter-edge 
scattering rate, i.e.~rate of backscattering from one edge of the Hall bar 
through the insulating bulk region to the other edge.
 
In the case of an antidot superlattice as studied here, not only the extended 
edge states at the boundaries of the hall bar exist, but also localized states 
around the antidots. We propose that backscattering in our superlattice takes 
place in several steps via the current loops localized around the antidots. 

We will now briefly discuss our data for different other filling factors 
(magnetic fields). At a filling factor of $\nu =1.9$ (B~=~7.2~T) 
(Fig.~\ref{0317UI}(b)) the cyclotron resonance behaves similiar to $\nu =2$ 
(B~=~6.8~T) (Fig.~\ref{0317UI}c). Magnetoplasmon amplitude and $\rho_{xx}$ are 
both almost zero for small bias-currents below 2$\mu$A and then increase in a 
way very similiar to each other. This is explained as follows: for low 
bias-currents below $2\mu A$ the sample is in an adiabatic state of 
conductance and only the cyclotron resonance absorption produces a 
photoresponse. For higher bias-currents bulk transport is beginning to be 
thermally activated, such that the sample is in a dissipative state now. 
Correspondingly also the magnetoplasmon absorption generates a photoresponse. 
At a magnetic field of B~=~6.2~T, still in the plateau region of $\rho_{xx}$ 
around integer filling factor $\nu = 2$ (not shown), the data are very 
similiar to those for $\nu =2$ (B~=~6.8~T) (Fig.~\ref{0317UI}c). The reason 
for this is, that -- up to bias-currents of $10\mu A$ -- the 2DEG is still in 
an adiabatic state ($\rho_{xx}$ is still vanishing). Thus only the CR 
absorption generates a photoresponse. For a filling factor of $\nu=1.6$ 
(B~=~8.6~T) (not shown), where $\rho_{xx}$ is nonzero the data resemble those 
for $\nu=1.7$ (B~=~8~T) (where $\rho_{xx} \neq 0$ as well) 
(Fig.~\ref{0317UI}(a)). This is because the 2DEG is in a dissipative state. 
Thus both CR and MP absorption lead to a photoresponse. 

Now there remains to be discussed {\em why} the magnetoplasmon is visible in 
the photoresponse in the dissipative transport regime only, but the cyclotron 
resonance occurs both in the dissipative and the adiabatic transport regimes, 
i.e. why only CR absorption is able to induce backscattering of edge-states.

The extent of the edge-states into the 2D-bulk is comparable to the edge 
depletion length, which is of the order of a few 100~nm. 

The characteristic coherence length $l$ of a collective excitation with a 
frequency/wave vector-dependence $\omega (q)$ is generally equal to 
$l=(d\omega /dq)\tau$ with the scattering time $\tau$. For the magnetoplasmon 
in our samples we have $l\approx 10\mu m$. This means that the magnetoplasmon 
excitation has its full strength only a distance $l\approx 10\mu m$ away from 
the sample boundaries, i.e. deep in the 2D-bulk. At the location of the edge 
states, not much more than 100~nm away from the boundaries, the magnetoplasmon 
has almost vanished. When only the edge-states contribute to electric 
conduction, as is the case under adiabatic transport conditions, the 
magnetoplasmon in the 2D-bulk cannot be converted into a photoresponse. This 
conversion could only be done by the edge-states, but at the location of the 
edge-states the magnetoplasmon oscillation strength is already negligible. 

The characteristic length for the cyclotron resonance is the magnetic length. 
Even though the cyclotron resonance as a collective phenomenon is affected by 
finite-size effects up to macroscopic dimensions \cite{Vasiliadou}, it is 
sensitive to the local potential landscape even in the nm-range \cite{Merkt}. 
Thus absorption in the cyclotron resonance can be converted into a 
photoresponse also by the edge states.
%---------------------------------------------------------------------

\section{Conclusion}

We have examined cyclotron resonance- and magnetoplasmon-induced changes in 
the longitudinal voltage of an antidot superlattice in high magnetic fields by 
means of photoconductivity measurements. In the dissipative transport regime 
we detect both magnetoplasmon and cyclotron resonances in the photoresponse. 
In the adiabatic transport regime at integer bulk filling factor $\nu = 2$ we 
detect only the cyclotron resonance in the photoresponse. To explain these 
experimental results, i.e. the different dependencies of the amplitudes of the 
photoresponse in the cyclotron resonance and in the magnetoplasmon resonance 
on the applied bias current at different filling factors, we suggest a model, 
wherein different mechanisms generate photoconductivity in the magnetoplasmon 
resonance and in the cyclotron resonance, respectively. According to this 
model electron heating is responsible for photoconductivity in the dissipative 
transport regime. In the regime of adiabatic transport at integer filling 
factor $\nu =2$ the photoresponse is generated by backscattering of the 
topmost edge state, and this backscattering can be caused by the cyclotron 
resonance only.

The reason for why only the CR can cause backscattering of edge states is that 
cyclotron and magnetoplasmon resonances have different characteristic lengths. 
The characteristic length of the cyclotron resonance is comparable to the 
lateral extent of the edge-states, thus CR absorption can induce 
backscattering of edge-states. The characteristic length of the magnetoplasmon 
is by almost an order of magnitude larger than the lateral extent of the 
edge-states, thus MP absorption can{\em not} induce backscattering of 
edge-states. 

\acknowledgments

We would like to acknowledge Alik Chaplik and Achim Wixforth for valuable 
discussions and the Deutsche Forschungsgemeinschaft for financial support. 

%---------------------------------------------------------------------

\begin{figure}
\caption{Sample layout and setup for photoconductivity measurements.}
\label{ADOTS}
\end{figure}

\begin{figure}
\caption{Photoresponse $\Delta U_{xx}$ at a bias current of $I=1~\mu A$ and a 
magnetic field of $B=8.4~T$ for (a) the antidot superlattice and (b) the 
unpatterned reference section. In the dashed part of curve (a) an artefact has 
been 
removed.}
\label{02131484T}
\end{figure}

\begin{figure}
\caption{Photoconductivity spectra at a magnetic field (filling factor) of (a) 
$B=8~T$ ($\nu =1.7$) and (b) $B=6.8~T$ ($\nu =2$) for different bias currents 
ranging from $I=-10~\mu A$ to $+10~\mu A$.}
\label{031780T}
\end{figure}

\begin{figure}
\caption{Amplitudes of cyclotron resonance (open circles) and magnetoplasmon 
(solid dots) photoresponse in comparison to the longitudinal resistance 
$dU_{xx}/dI_{bias}$ as a function of the bias current applied between source 
and drain.}
\label{0317UI}
\end{figure}

\end{document}